[1994/12/01]
\documentclass[draft]{article}
\title{ Size Effects in Heavy Ions Fragmentation}
\author{   A.  Barra\~n\'on  
\footnote{ Universidad Aut\'onoma Metropolitana. Unidad Azcapotzalco.
Av. San Pablo 124, Col. Reynosa-Tamaulipas, Mexico City. email: bca@correo.azc.uam.mx } ;
 J. A.  L\'opez
\footnote{Dept. of Physics, The University of Texas at El Paso. El Paso, TX, 79968  }    ;
C. Dorso.
\footnote{ Dep. de F\'{\i}sica, Universidad de Buenos Aires. Buenos Aires, Argentina}   }

\date{October, 20th, 2002}

\usepackage{amsmath,amsthm}

\chardef\bslash=`\\ 
\newcommand{\ntt}{\normalfont\ttfamily}

\theoremstyle{definition}

\theoremstyle{remark}

\newcommand{\eval}[2][\right]{\relax
  \ifx#1\right\relax \left.\fi#2#1\rvert}

\newcommand{\envert}[1]{\left\lvert#1\right\rvert}

\begin{document}
\maketitle
\markboth{Sample paper for the {\protect\ntt\lowercase{amsmath}} package}
{Sample paper for the {\protect\ntt\lowercase{amsmath}} package}
\renewcommand{\sectionmark}[1]{}

\abstract
Rise-Plateau Caloric curves for different Heavy Ion collisions have been obtained, in the range of experimental observations. Limit temperature decreases when the 
residual size is increased, in agreement with recent theoretical analysis of 
experimental results reported by other Collaborations. Besides, promptly emitted 
particles influence on temperature plateau is shown. LATINO binary interaction 
semiclassical model is used to reproduce the inter-nucleonic forces via Pandharipande
 Potential and fragments are detected with an Early Cluster Recognition Algorithm.

\section{ Introduction}

M. Rousseau et. al. have studied light charged particle emission from a
compound nucleus of $^{40}Ca$ obtained from $^{28}Si$+$^{12}C$ reaction, 
providing evidence of binary emission of $^{8}Be$,
 in a data fitting scenario formed by a sequential evaporation process, 
a subsequent fragment emission from the $^{40}Ca$ compound nucleus and finally  
the residual nucleus freeze-out \cite{Rousseau}.

In the case of inverse kinetic reactions, fragmentation occurs in two stages,
namely a first pre-equilibrium stage where participants interact and deposit excitation 
energy in the spectators, and a second stage where the excited spectator residual decays
 \cite{Bauer}.

Considering small excitations, there is a delay in fission fragment emission in 
the range of  1000 fm/c  and for larger excitations the charged particles are 
emitted at the same time as heavy fragments ($Z_{imf}$= 8- 16), with an integral 
rupture time in the range of 200 fm/c after thermalization \cite{Pienkowski}.
 Kolomietz et. al. have obtained ``rise-plateau-rise isobaric caloric'' curves in the 
Thomas-Fermi approximation with an effective force SkM, showing that in the plateau 
there is liquid-gas coexistence  \cite{Kolomietz}.

 Gourio et. al. have obtained light particle emission times around 200 fm/c, 
using two-particle correlation functions in order to analyze INDRA and GANIL data
\cite{Gourio}. D'Enterr\'{\i}a et. al. have obtained similar caloric curves 
studying Brehmstahlung photon emission in Heavy Ion Collisions using a thermal model
 \cite{Enterria}.
 D'Agostino et. al. have shown the usefulness of an abnormal change in kinetic energy 
variance as a first order phase transition signature \cite{Agostino}. Nuclear collisions 
for energies in tenths of $meVs/nucleon$ produce excited systems that break-up into 
several intermediate size fragments \cite{1}. These multifragmentation phenomena, similar to those occurring in inertial confinement reactions \cite{2} and in the nanostructures surface synthesis \cite{3}, could be the adaptation of a macroscopic phase transition to a finite and transient system.

In the realm of nuclear physics, the possibility of reaching a critical behavior 
in heavy ions collisions has motivated several studies on critical exponents. 
The first one focused on proton-Xe and proton-Ar collisions by Purdue group
 \cite{4, 5} and others afterwards \cite{6}. More recently, modern detection 
technology allowed caloric curve experimental computation \cite{7, 8} , 
i.e. the relation between system temperature and excitation energy at  
fragmentation time. Nevertheless, the influence of residual finite size, 
a transient evolution and experimental limitations due to observing only final fragments, turn out to be problematic.

As explained by Pochodzalla y Trautmann \cite{9} , 
there are difficulties in the reconstruction of laid down energy via 
exit channels and therefore in caloric curve computation. Other problems 
arise from the fluctuations in system size \cite{10}, 
``side feeding'' effects on final mass distribution, and in the use of final spectra modified by collision evolution \cite{12}.

As a matter of fact, experiments using different ``thermometers" 
have reached contradictory caloric curves such as the typical 
rise-plateau-rise plot \cite{13}, a peculiar ``rise-plateau" caloric curve \cite{14}, and
 even a ``rise-rise" caloric curve without the plateau characteristic of a first-order phase transition \cite{15}. 

The suppression of the final temperature growth could be related either to a low density rupture \cite{16}, or to a high fragmentation energy  destroying hot intermediate fragments \cite{9}.

With so many variables affecting the nuclear caloric curve determination, 
the question to answer is whether it is possible to extract this caloric curve in nuclear collisions. More recently, this was dealt with a Molecular Dynamics study in nuclear systems inside of a container \cite{17} and for a freely expanding classical system  \cite{18}.

Even when those computations can not be compared, neither between them nor with 
experiments, those exercises were able to obtain a caloric curve. Sugawa and
 Horiuchi \cite{17}, used Antisymmetrized Molecular Dynamics, 
in order to study a uniformly excited, contained system with constant pressure,
 and obtained a ``rise-plateau-rise" caloric curve. On the other side, Strachan and Dorso \cite{18} used a uniformly excited Lennard-Jones system with free expansion and  obtained a caloric curve with a ``rise-plateau" shape. The difference between these results 
arises from the collective expansion manifest in finite systems 
(but absent in infinite and contained systems), acting as an energy reservoir and avoiding a final temperature rise \cite{18}.

And Gobet et. al. have obtained event by event ``rise-plateau-rise" caloric
 curves for collisions of hydrogen clusters ions  with Helium targets, considering this caloric curve as a first-order phase transition signature in a finite system. The 
plateau is interpreted as latent heat associated with this phase transition \cite{Gobet}.

Notwithstanding their differences, these exercises suggest that fragmenting small systems may provide information about the caloric curve. Nevertheless, in 
regards of the fundamental role played by the geometry of the fragmenting system, i.e. contained systems versus free-expansions systems, a subtle question might be 
that of the role played by the collision in caloric curve computations. 
Namely, the influence of induced large correlations due to non linear dynamics
 on the obstacles found to compute the caloric curve.

   In this study a computation is performed that reproduces experimental 
collisions in order to obtain the caloric curve and examine the dependence 
of the limit temperature on the residual size. 

We use the weaponry developed in \cite{18} to focus on colliding excited systems in order to obtain the caloric curve and its dependence 
on the residual size. The manuscript is organized in the following way: after describing the model used, the fragment recognition algorithm and the persistence in section II, Section III establishes an effective way to compute the collision dynamic stages and section IV studies the temperature in the fragmenting system 
participant region. Caloric curve dependence on residual size is shown in section V, and some final conclusions close the article in Section VI. 

\section{ Latino Model.}

LATINO Model \cite{Barra99} uses the semi-classical approximation to simulate Heavy Ions collisions via a binary interaction Pandharipande potential. This potential is formed by a linear combination of Yukawa potentials, with the coefficients 
fixed in order to reproduce the properties of nuclear matter. Clusters are detected 
with an Early Cluster Recognition Algorithm that optimizes the configurations in 
energy space. Ground states are produced generating a random configuration in 
phase space, gradually reducing the velocities of the particles confined in a parabolic 
potential, until the theoretical binding energy is reached.

Due to obscurities in the break up process, a complex model is needed with 
the auxiliary tools required. Collision evolution is modeled with a
Molecular Dynamics code, though since Molecular Dynamics operates at a
nucleonic level, it is necessary to transform particle information in terms 
of fragment information via a fragment recognition algorithm. Using caloric curve 
as a phase transition signature, requires the knowledge of the break up time, using 
a property known as partition ``persistence". 

   Molecular Dynamics advantages to study nuclear collisions have been established previously.
 In this work we study the time evolution of central collisions numerical simulation. Target
 consists of a 3D droplet in its ``ground state". The projectile is a randomly oriented droplet
 in its ``ground state", boosted on the target with different kinetic energies. 

   Numerical integration of the equations of motion is performed through a Verlet algorithm at
 time intervals that ensure energy conservation at least in a 0.05 \%. Fig. 1 shows the evolution
 of a typical experiment.

The range for the projectile kinetic energy starts from $E_{beam}= 800$  $MeV$ in the center of
 mass reference system with about two hundred collisions for each projectile energy. Covering a
 wide energy range, we obtain different dynamical evolutions: starting from events where the 
projectile is absorbed by the droplet surface up to events with an exponential
 decay mass spectra. 
As shown in Fig. 2, the shapes of the spectra converge to a power law in this range of 
intermediate energies. The temperature of the participant region is computed using Kinetic Gas
 Theory and system excitation is calculated with the energy deposited in the residual. When the
 projectile energy is increased, the multiplicity is also increased as the system enters into the
 phase coexistence region (Fig. 3b) 
\cite{Barra01}.

With the fragments detected in phase-space, Mean Velocity Transfer is computed in the following way :
\begin{equation}
MVT_i= \sum_k \envert{v_{k,i} (t+dt) - v_{k,i} (t)}
\end{equation}

\subsection{ Fragment Recognition}

In order to obtain information about the fragments, Molecular Dynamics microscopic data must be analyzed through a  Fragment Recognition Algorithm. Although there are many, we will now describe the one used here.

In the ``Minimum Spanning Tree in Energy'' algorithm (MSTE), 
a given set of particles i,j,...,k belongs to the same cluster $C_i$ if: $ \forall i \in C_i,$  exists 
$j \in C_i / e_{ij} \le 0$ where $e_{ij}= V(r_{ij})+ ( \mathbf p_{i}- \mathbf p_{j})^2 / 4 \mu $, and
  $ \mu $ is the reduced mass of the couple {i,j}. MSTE searches configuration correlations between particles, 
considering their relative moment. Due to its sensitivity to recognize early particles, it is
 extremely useful to study the pre-equilibrium energy distribution of the participant particles.

\subsection{ Partition Persistence}

In order to use the caloric curve as a phase transition signature, 
both system excitation and temperature at phase transition are needed. 
Hence, break-up time is required, and is calculated in terms of 
the fragmentation time, $\tau_{ff}$,  namely a time when fragments evaporate 
only some monomers.  $\tau_{ff}$ can be detected comparing partitions 
at different times, which can be done using the ``Microscopic Persistence Coefficient'',
 $P$ \cite{23}:

\begin{equation}  P \left[ X,Y \right] = 
{\frac {1}{ \sum_{cl} n_{i}} }     \sum_{cl} n_{i} \frac{a_{i}}{b_{i}}
\end{equation}  
where $ X \equiv \{ C_i \}$ and $Y \equiv \{ C'_i \}$ 
are different partitions,
$b_i$ is the number of particle pairs in the cluster
$C_i$ of partition $X$, $a_i$ is the number of particle pairs belonging to cluster
 $C_i$ that remain paired in a given cluster $C'_j$ 
of partition $Y$, $n_i$ is the number of particles in cluster $C_i$. 

 $P \left[ X,Y \right]$ is equal to 1 if the microscopic
 composition of partition $X$ is equal to that of partition $Y$, 
and tends to 0 when none of the particles belonging to a given cluster $X$ 
remain together in any cluster of $Y$.
 
It is convenient to study the time evolution  for the quantities:
\begin{equation} \hat P^{+} \left[ X(t) \right] \equiv
\left\langle P \left[ X(t), X(t \to \infty ) \right] 
\right\rangle_{events}
\end{equation}
\begin{equation}
 \hat P^{-} \left[ X(t) \right] \equiv 
\left\langle P \left[ X(t \to \infty ), X(t) \right] 
\right\rangle_{events}
\end{equation}
\begin{equation}
 \hat P^{dt} \left[ X(t) \right] \equiv
\left\langle P \left[ X(t), X(t+dt ) \right] 
\right\rangle_{events}
\end{equation}
where $X(t)$ represents a partition computed at time t, $X(t \to \infty)$
is an asymptotic partition, and  $\left\langle \right\rangle_{events}$ 
represents an average on the total set of collisions. 
In simple terms, 
$\hat P^{+} \left[ X(t) \right] $ 
determines whether all particles bound at time t remain bound asymptotically.

In a similar fashion $\hat P^{-} \left[ X(t) \right] $ measures the reciprocal value
, $i.e.$, the degree in which an asymptotical partition is contained in a partition
 occurring at time t. In conjunction, $\hat P^{+}$ y $\hat P^{-}$
can be used to analyze the evolution of the microscopic composition of the
most bound partition, until it reaches its asymptotic form.

   Finally $\hat P^{dt}\left[ X(t) \right] $ provides information on 
the given partition $activity$, and can be thus used to define 
the fragmentation time, $t_{ff}$, once it reaches some degree of stabilization.

It is useful to work with some $normalized$ quantities:
\begin{equation}
P^{+} \left[ X(t) \right] = 
 \hat P^{+} \left[ X(t) \right] 
/ \left\langle P \left[ X(t\to \infty), X^{'}(t \to \infty ) \right] 
\right\rangle_{events} 
\end{equation}
\begin{equation}
P^{-} \left[ X(t) \right] = \hat P^{-} \left[ X(t) \right] 
/ \left\langle P \left[ X(t), X^{'}(t ) \right] \right\rangle_{events}
\end{equation}
\begin{equation}
P^{dt} \left[ X(t) \right] = \hat P^{dt} \left[ X(t) \right] 
/ \left\langle P \left[ X(t+dt), X^{'}(t+dt ) \right] \right\rangle_{events} 
\end{equation}
where $X'$ is the partition formed by the clusters belonging to partition
 $X$, and where each fragment has evaporated  a particle.
Normalized quantities compare real values of $\hat P's$ 
with an evaporative level of reference.
 $P^+$ , $P^-$ and $P^{dt}$, refer to distinct persistence coefficients, namely forwards,
 backwards and differential, respectively.

\section{Dynamic Evolution}

Armed with the tools introduced before (MD, MSTE and $P^+$ 
, $P^-$ and $P_{dt}$) we proceed to study the dynamical evolution of 
the collisions described in section II. Analyzing the Mean Velocity 
Transfer, we can characterize the collision stages and identify  
 early emitted particles. This allows to study the excitation energy 
and the detection of the fragmentation time.

\subsection*{A   Collision Stages}

Two colliding stages are observed, with an initial highly colliding stage
 produced when the projectile hits the droplet surface and the energy is distributed chaotically. Collisions at this initial stage form a shock wave responsible for the prompt emission of light energetic particles from the surface. As the 
shock wave travels into the droplet, it produces density fluctuations and internal fractures as a consequence of the momentum transferred and initiates disordered collisions leading to an excitation thermalization. 

   In order to reach a deeper understanding of this process, we compute the 
``Mean Velocity Transfer'', defined by:
\begin{equation}
 M_j (t) =  \left\langle  \sum _{i=1}^{N} \envert{ ( \vec v_{i} (t + dt) - 
\vec v_{i} (t) ) \cdot \hat e_j }  \right\rangle_{events} 
\end{equation}
where j denotes incident and normal directions and $\hat e_j$ 
is a unitary vector in these directions. Fig. 5
shows time evolution of $M_x$ for distinct values of $E_{beam}$ 
for Ni+Ag central collision.

Isotropic collisions ($disordered$ $colliding$ $mode$ ) are responsible 
of the momentum redistribution among the particles remaining in the system. This energy heats the system and constitutes an expanding collective motion. The disordered colliding mode is the only one present for uniformly excited systems where it is responsible of fragmentation and its outwards flow disperses the fragments \cite{18, 20}.

MSTE algorithm can be used to study the size of the MSTE biggest fragment, total multiplicity and persistence coefficients. Figs. 3a, b and c show time evolution of these three quantities. The description suggested is that, due to an initial violent collision, some particles acquire enough energy to be released of the vicinity or interior of a cluster. This reduces the mass of the biggest MSTE cluster  in this early stage and increases the total multiplicity. This tendency is maintained until the mean momentum for each particle 
permits the binding of particles 
configurationally close. After the initial reduction of the initial biggest fragment
 size, a coalescent behavior promotes the multiplicity decrease and the biggest 
cluster size grows until it reaches 
a maximum. This time signs the end of the initial energy distribution and can be used to define the pre-equilibrium time.

   Persistence coefficients can be used to understand how the partition reaches its asymptotic microscopic composition. Fig. 3c shows the time evolution of $P^{+} \left[ X(t) \right] $  and $P^{-} \left[ X(t) \right] $ computed using the MSTE partitions for a $E_{beam}$ . 

Backwards Persistence Coefficient shows initially a decrease due to 
the fragment formation stage, followed by a subsequent increase due to 
the dynamic re-absorption lasting until $t_{pre}$. High values of $P^-$ indicate that more particles belonging to a given cluster 
remain paired at time t. Consistently, $P^+$ coefficient shows a plateau during this re-absorption stage that is extended until time $t_{pre}$,
 followed by a monotonic increase due to an evaporative dynamics of the MSTE clusters. 
 During this stage, MSTE algorithm detects a large biggest cluster, revealing that 
the system is still dense.

In summary, an initial stage characterized by the existence of a shock wave releasing particles from the surface, the reduction of the biggest fragment size and an increase of the multiplicity ($i.e.$ decreases $P^-$). Afterwards, the shock wave  crosses the droplet distributing uniformly the energy ($M_x \sim M_y$) and a coalescent behavior is installed signed by the increase of  $P^-$ and deleting the memory of the entrance channel, reducing the multiplicity and augmenting the size of the biggest fragment until it reaches a $\sim 80 \% - 90 \%$ of the total mass. This is followed by an evaporative dynamics of the clusters as indicated by a monotonic increase of $P^+$.  
    Fig. 4 shows colliding stages for Ni+Ag with an energy equal to 1600 MeV. 
A first stage ends with a peak in the kinetic energy transported by light particles, signed by a peak in the black curve. A second stage ends with the attenuation of the intermediate fragments production (pink curve). Between these peaks, a peak in Mean Velocity Transfer  
(orange curve) is observed. Hence, Mean Velocity Transfer (MVT) produces  instabilities leading to light particles emission and subsequent intermediate fragment emission. Once the MVT is stabilized, a final stage follows  characterized by the emission of light fragments and system freeze-out.

\section{Temperature}

The next step to obtain a caloric curve is the computation of the system 
temperature at fragmentation time, using the kinetic energy of nucleons in the participant region, relative to the center of mass.
 This seems to be justified as long as the excited droplets reach thermal 
equilibrium at break-up \cite{18}. Nevertheless, there exist objections  
with respect to the use of thermodynamic concepts in small finite  
transient systems \cite{24,25}. 

Now we define this temperature and study its time evolution.

\subsection*{A      Participant Region Temperature.}

   In central nuclear collisions, shock waves orthogonal to the beam direction arise, separating compressed nuclear matter from cold nuclear matter. As the projectile penetrates into the target, a region with both high density and excitation is formed. The complex collectively expands in a direction normal to the beam due to the fact that pressure is null in the exterior of the system and therefore expansion in  beam direction is decreased.

A discontinuity appears in the contact point between nuclei, with both nuclei traveling in opposite directions with respect to the center of 
mass, that evolves into two shock waves traveling in opposite directions 
across the projectile and the target \cite{Danielewicza}.   

A study based on the dynamic model of several fluids, showed that 
most of the energy associated with the stopping derived from transverse  collective motion, is used for mid-rapidity fragment production  \cite{Dimitrua}. In the contact region between both nuclei, frictional forces promote energy dissipation and particle deflection, since projectile participant nucleons interact several times with target participant nucleons. \cite{Wilczinskia}. All along this discontinuity, transport effects are expected leading to a soft change in the properties instead of a discontinuous transition \cite{Mornasa}.

 In this participant region a temperature can be obtained, considering the
 nucleons in this region and computing the temperature provided by 
the kinetic energy of these nucleons measured with respect to the center of mass. 
 Promptly emitted, not reabsorbed nucleons are excluded, using the MSTE 
Early Cluster Recognition Algorithm for this sake.

\subsection*{B       PEPS and Energy Excitation.}

As it happens in the experimental case, the observed presence of promptly emitted particles(PEP) makes possible the $a$ $priori$ computation of the fraction of beam energy transformed into excitation energy. This complication is absent in computations for both contained or infinite systems, and requires a sensible definition of the excitation energy that should depend on the PEPs.

PEPs can be defined as unbound light clusters (with mass $\le$ 4) detected by the algorithm at time $t_{pre}$ and that remain unbound at any posterior time ($i.e.$ no re-absorption). Once PEPs are defined, their kinetic energy can be used to estimate the energy remaining in the system.

Figs. 6a and 6b show that the transported energy by PEPs and the number of PEPs as a function of the beam energy, respectively. With the number of PEPs and their energy quantified, the energy remaining in the system after time $t_{pre}$ can be also obtained.

Fig. 6c shows that energy is deposited in the target as a function of the beam energy. These figures show that the fraction of available energy leaving the system, as a consequence of this prompt emission, is considerable. Even more, the energy remaining in the system shows a saturation behavior indicating that there exists a limit for the quantity of energy that can be transferred in a collision, which does 
not occur in uniformly excited systems.

\subsection*{C     Fragmentation Time}

After the $pre$-$equilibrium$ time, $t_{pre}$ (cf. section IIIA), 
the energy remaining in the system is completely distributed, and the density correlations $\bf r- \bf p$ induced by the initial shock wave begin to constitute a collective expansion. Eventually, this expansion will transform the initial fluctuations into fragments well defined in space $\bf r$, recognizable for MSTE algorithm. Caloric curves should reflect the system state along the phase transition, and we consider this time as the fragmentation time, $t_{ff}$, associated with the stabilization of MSTE density fluctuations

\section{Caloric Curves and Residual Size}

Once the collision dynamic simulation is performed, prompt fragments are detected, fragmentation times are identified and excitation energy as well as temperature are computed, a caloric curve can be obtained. Formally, this quantity investigated experimentally \cite{7, 8, 13, 14, 15, 26} as well as computationally \cite{17, 18, 22}, is the functional relation between the temperature of the system and its excitation energy. In this study we
 extend this analysis to Pandharipande droplets  excited by collisions.

  Figs. 7-9 show the caloric curves computed for a wide energy range 
of excited collision systems built from participant region temperatures at time $t_{ff}$. As can be seen, caloric curves are similar to those obtained from uniformly excited systems \cite{22}. Once again, the relevant characteristic is the almost constant temperature behavior in fragmentation region. In other words, collision data provide a ``rise-plateau" caloric curve. Besides, these caloric curves portray a limit temperature that diminishes as the residual size 
increases.
 
\section{Conclusions.}

Event by event ``Rise plateau'' Caloric curves have been obtained for different residual
 size bins for Ag+Ag Central Heavy Ion Collision. Temperature is computed using Kinetic
 Gas Theory in the Participant Region, in the moment when the persistence reaches a
 value close to one. When the residual size increases, a clear decrease of the limit
 temperature is observed. This is in agreement with recent studies of experimental
 data performed by Natowitz et.al. \cite{Natowitz}. 

 Therefore we have confirmed using a dynamical model, recent analysis of 
experimental data, where a decrease of the limit temperature is observed as 
the residual size is increased.

These results motivate us to perform future studies on the influence of the dynamics 
in the formation of a plateau characteristic of the coexistence region, for different
 residual sizes. Authors acknowledge financial support from NSF through PHYS-96-00038
 fund and free access to the computational facilities of The U. of Texas at
 El Paso and UAM-A.

\end{document}